\documentclass[5p]{elsarticle}

\usepackage{lineno,hyperref}
\modulolinenumbers[5]

\journal{Arxiv Preprint}
\usepackage{amsmath}
\usepackage{amssymb}
\usepackage{float}
\let\textquotedbl="

\begin{document}

\begin{frontmatter}

\title{In-vitro Major Arterial Cardiovascular Simulator to generate Benchmark Data Sets for in-silico Model Validation}

\author[thmaddress]{Michelle Wisotzki}

\author[thmaddress]{Alexander Mair}
\author[tuaddress]{Paul Schlett}
\author[thmaddress]{Bernhard Lindner}
\author[thmaddress]{Max Oberhardt}
\author[thmaddress,tuaddress]{Stefan Bernhard\corref{mycorrespondingauthor}}
\cortext[mycorrespondingauthor]{Corresponding author}\ead{stefan.bernhard@lse.thm.de}

\address[thmaddress]{Technische Hochschule Mittelhessen - University of Applied Sciences,  Life Science Engineering,  Gießen,  Germany}
\address[tuaddress]{Freie Universität Berlin, Institute of Mathematics, Berlin, Germany}

\begin{abstract}
A deeper understanding of the influence of common cardiovascular diseases like stenosis, aneurysm or atherosclerosis on the circulatory mechanism is required, to establish new methods for early diagnosis. Different types of simulators were developed in the past to simulate healthy and pathological conditions of blood flow, often based on computational models, which allow to generate large data sets. However, since computational models often lack some aspects of real world data, hardware simulators are used to close this gap and generate data for model validation.\\
The aim of this study is the development and validation of a hardware simulator to generate benchmark data sets of healthy and pathological conditions. The in-vitro hardware simulator in this study includes the major 33 arteries and is driven by a ventricular assist device generating a parametrised input condition at the heart node. Physiologic flow conditions including heart rate, systolic/diastolic pressure, peripheral resistance and compliance are adjustable in a wide range. The pressure and flow waves at 17+1 different locations are measured by inverted fluid resistant pressure transducers and one ultrasound flow transducer supporting a detailed analysis of the measurement data. The pressure and flow waves show physiological conditions. Furthermore, the influence of stenoses degree and location on blood pressure and flow was investigated. The results indicate decreasing translesional pressure and flow with increasing degree of stenosis, as expected.\\
The benchmark data set is made available to the research community, with the purpose to validate and compare in-silico models of different type. 

\end{abstract}


\end{frontmatter}

%\linenumbers

\section{Introduction}
The prevalence of cardiovascular diseases is increasing worldwide \citep{Fowkes.2013}. Commonly atherosclerosis, stenosis and aneurysms are the major reason. Mortality is increasing with age and is also dependent on gender \citep{Mathiesen.2001}.
Early diagnoses of these diseases are desirable, consequently a deeper understanding of the influence of arterial diseases on the underlying system morphology and flow properties is necessary. Besides imaging techniques, which are often expensive and not available at primary physician level, there are currently no suitable mass screening methods to assess specific arterial properties at required accuracy. However, continuous quantities, like the Photoplethysmogramm (PPG) or pressure and flow, are obtained easily and contain information about the vascular structure, thus it would be desirable to infer the arterial properties from these signals \citep{Quick.2001}. 

Nowadays, on the other side, a variety of in-silico simulation models were developed to gain a deeper understanding of the circulatory mechanism by simulating healthy and pathologic conditions of cardiovascular blood pressure and flow by means of numerical models \citep{Huttary.2017, Gul.2015, Quarteroni.2016, Quarteroni.2001, Zenker.2007, Garber.2021, Jezek.2017}. 
Given the patient-specific morphology and parameters, these computational simulation models are able to generate large data sets for the state variables of pressure and flow. In \citep{Jones.2021} for e.g., a virtual patient database was generated to study the influence of arterial diseases on the hemodynamics, by using a detailed arterial network from \citep{Boileau.2015}. 
In \citep{Huttary.2017}, a confrontation of aorta (CoA) was simulated and has been successfully used to setup and identify patient-specific models and to reconstruct pre- and post-treatment scenarios characterized by available routine clinical data sets. The authors state that, for accurate remodelling of clinical situations, data acquirement in the clinic has to become more reliable to keep uncertainties small. 
Furthermore, due to the simplified model complexity (e.g. dimension reduction, shape optimisation, linearisation, etc.) data sets of numerical simulations lack some aspects of the real world data of the corresponding cardiovascular system. Consequently, such models have not yet made their way to clinical routine, because validation is still problematic \citep{VignonClementel.2022}.

Hardware simulators try to close this gap by generating parametric data sets of pressure and flow for model validation. In the last decade different types of in-vitro hardware simulators of the cardiovascular system were developed, mainly to verify computational fluid dynamics models \citep{Jin.2021, Korzeniowski.2018}, to understand specific fluid dynamical conditions \citep{Gehron.2019}, or to validate ventricular assist devices \citep{Ferrari.2002, Pugovkin.2015, Pugovkin.62920176302017}.
In \citep{Gehron.2019}, a life-sized mock circulatory loop of the human circulation was developed for fluid-mechanical studies using an extracorporeal life support system and two pneumatically driven vascular assist devices (VADs) representing the left and right ventricle.  Furthermore, mock circulatory loops often include no detailed mapping of the arterial system for test and validation of ventricular assist devices \citep{Ferrari.2002}. However, in \citep{Jin.2021} waveform measurements in a silicone arterial tree are compared to numerical predictions of a visco-elastic 1-D model to test the accuracy of the non-linear 1-D equations of blood flow in large arteries. 

However, none of the hardware simulation setups was used as a tool to generate data sets containing relevant information about specific diseases for diagnostic purposes. Thus, the aim of this study is the development and validation of a patient-specific cardiovascular simulator to generate parametrical data sets, facing benchmark problems that characterize for e.g. the influence of arterial stenosis within the cardiovascular system and make these data sets available to the research community. Therefore, a Major Arterial Cardiovascular Simulator (MACSim) was developed and extended over the past years, integrating patho-physiological  information to improve the  understanding and validity of computer simulation models for interpretation in a clinical setting. 

Within this work, a arterial network of the 33 major arteries was realised, the corresponding vessel morphology and parameters are presented. Furthermore, a detailed description of the measurement setup and procedure, including the definition and explanation of the different measurement scenarios, is given.

The physiologic measurement scenarios in this work were defined to quantify the impact of arterial abnormalities (e.g. stenosis) on the pressure and flow waves within the circulatory system. The pathological conditions of stenosis with different degree and location were addressed. Generated data sets are designed for the validation of computational simulation models to enable a community wide comparable statement of their quality. Specific data sets could be generated on author request. Moreover, the calibration of the pressure and flow sensors was established with high accuracy to allow high grade model validation. Finally, the measurement results of the different measurement scenarios are presented and discussed.

\section{Materials and Methods}

\subsection{Cardiovascular Simulator}
The development process of the simulator was led by six main design criteria with the aim to establish a modular and flexible simulation environment that is able to produce large statistical data sets of specific diseases within highly reproducible flow conditions:

\begin{enumerate}
\item Minimization of the pulse wave reflection with the condition to obtain realistic wave reflections from peripheral bifurcations and pathologies.
\item Adjustable flow conditions to a wide range of physiological conditions like for e.g. heart rate, systolic pressure, compliance, peripheral resistances, etc.
\item Measurement of pressure and flow at several different locations within the cardiovascular simulator.
\item Improved laboratory conditions for a highly reproducible pressure and flow measurement on sample a accurate time basis.
\item Parametric scripting of ventricular boundary conditions.
\item Persistent data management in relational data base for post-processing.
\end{enumerate}

The multivariate statistical data sets include relevant meta-information about the experiments and are stored to a MySQL database for further analysis. In the context of this study the data set is made available via Matlab files for simple community wide post-processing. MySQL data can be obtained on author request.\\

\begin{figure}%[H]
\centering
\includegraphics[width=\linewidth]{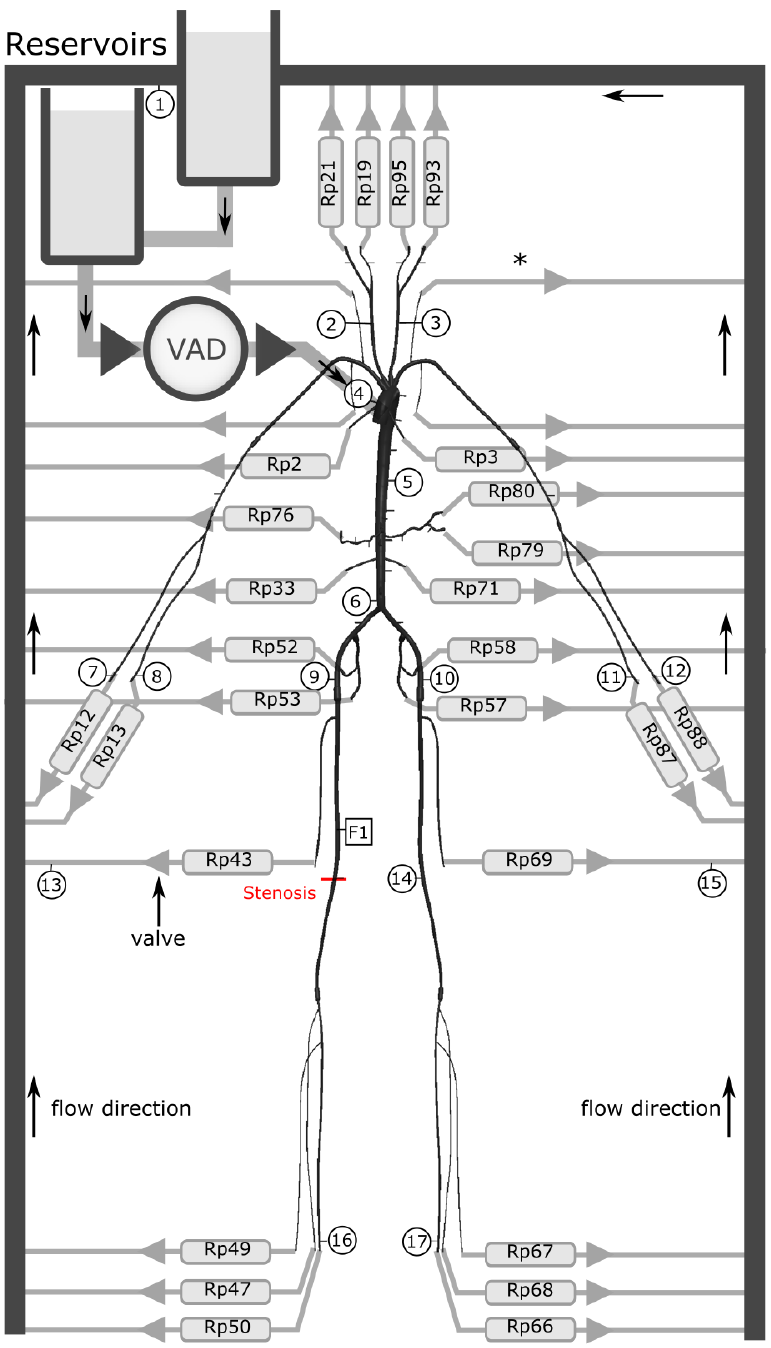}
\caption{Schematic  of the experimental setup including a VAD pump and the vascular model. The resistance elements (grey boxes) with adjacent check valves separate the arterial and venous section. A water-glycerine mixture (approx. 60/40 weight \%) of viscosity $\eta = (3,88 \pm 0,1) \, mPa \cdot s$ was used to model the properties of blood. F1 and 1 to 17 represent the measurement locations of the flow and pressure sensors, respectively. The compliance elements (syringes, see figure \ref{PeripheralElements}) are located at the peripheral ends (prior the peripheral resistances Rp) and at the * marked position, except for Rp52, Rp53, Rp58 and Rp57.}
\label{schematic}
\end{figure}

The experimental setup of the simulator consists of the following main components (see figure \ref{schematic}): arterial and venous system with valves and reservoirs, heart pump, compliance and peripheral resistance elements and pressure and flow sensors.

\begin{figure*}%[b] fuer untere Anordnung
\centering
\includegraphics[width=\linewidth]{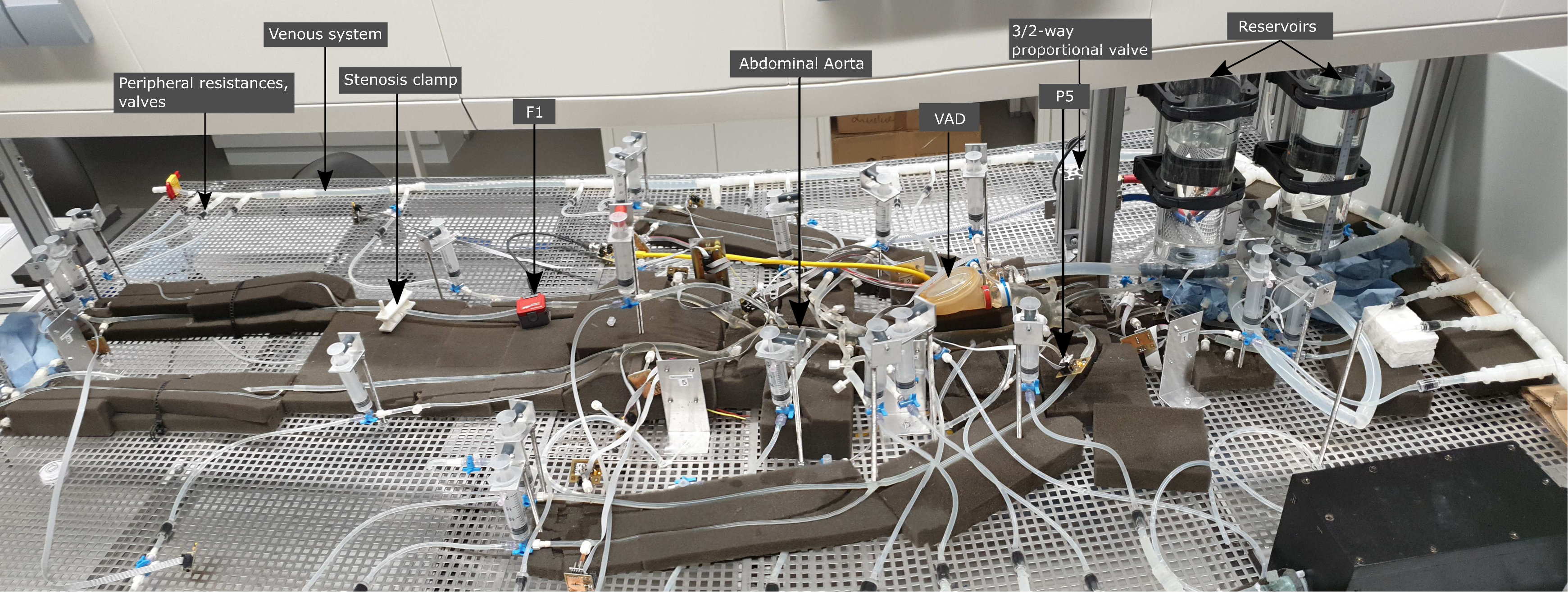}
\vspace{-3mm}Ftrapezoid

\caption{Major Arterial Cardiovascular Simulator (MACSim) including the main components: 3/2-way proportional valve, VAD, arterial system, venous system and reservoirs, peripheral resistances, valves, etc.}
\label{macsim}
\end{figure*}

\paragraph{Arterial and Venous System}
The structure of the cardiovascular simulator contains the major 33 arteries down to an inner diameter of 1,5 mm. This artificial arterial system is realised by a system of silicone tubes, which have similar characteristics such as inner and outer diameter, length and elasticity of the corresponding human arteries. The structural data for the arterial network was obtained from a real patient-specific MRI scan, followed by simplification and smoothing of the boundaries. Thereby, the individual parts of the arterial vascular system (aorta, art. carotis, art. subclavia, art. celiaca, art. iliaca and art. femoralis) were fabricated and assembled using injection molding. The other parts of the arterial system were made from standard silicon tubes due to the low vessel complexity and diameter. The whole vascular system is bedded on individually shaped PU-foam blocks, to ensure a proper anatomical tethering.
In addition to the arterial vascular system, the simulator includes a venous return system and two reservoirs connecting the venous and arterial system (see figure \ref{schematic} and \ref{macsim}).

Since only measurements on the arterial vascular system are performed, a detailed mapping of the venous system was omitted and instead a simple feedback from the individual peripheral arteries to the reservoirs was realised. Both reservoirs are filled with $V_R = 985 \, ml$ of fluid, thus creating a hydrostatic pressure offset $p_h = 14,42 \, mmHg$ throughout the model cardiovascular system. The systems diastolic blood pressure, \v{p}, is set by a combination of the peripheral flow resistances, $R_p$, and the level in the reservoirs. 

The viscosity and density of the fluid in the simulator are adjusted to achieve physiological values for human blood by a water-glycerine mixture (approx. 60/40 weight \%) , i.e. a resulting density of $\rho = (1,094 \, \pm 0,002) \, g/ml$ and a dynamic viscosity of $\eta = (3,88 \pm 0,1) \, mPa \cdot s$ at room temperature $\vartheta = 22,4 \, ^\circ C$.

The node numbering of the arterial network refers to the computational simulation modelling environment SISCA \citep{Huttary.2017}. In this software framework for multi-compartment lumped modelling each peripheral node number (see appendix figure \ref{sisca}) represents a flow resistance Rp in figure \ref{schematic}. The corresponding table \ref{structuralMeasures} contains measurements and estimations for the vessel diameter d, length l, wall thickness h and elastic modulus E.

\begin{figure} [H]
\centering
\includegraphics[width=\linewidth]{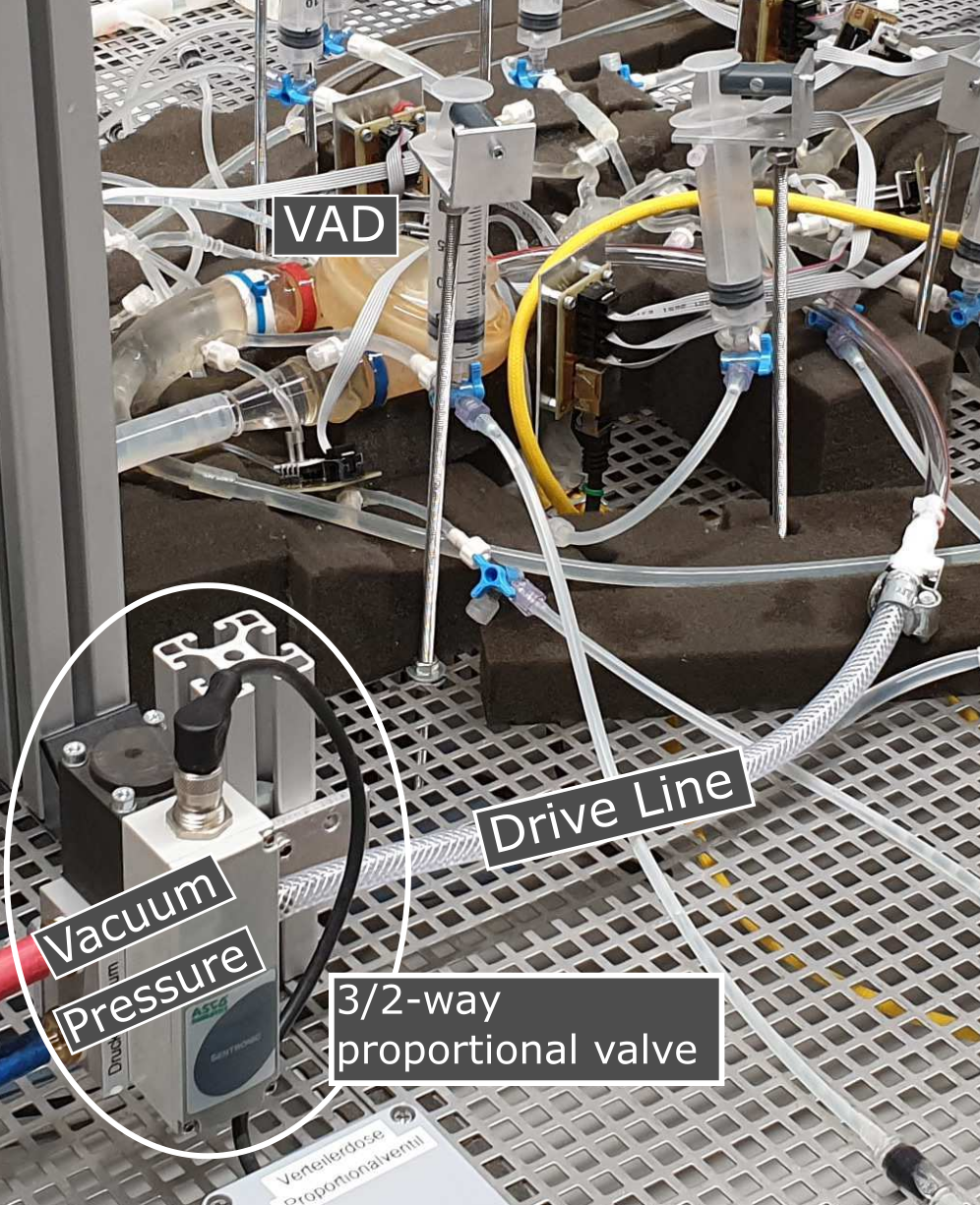}
\vspace{-3mm}
\caption{3/2-way proportional valve and the VAD. The 3/2-way proportional valve mixes relative underpressure of $ p_u = 0,4-0,7 \, bar$ and overpressure of $p_o = 1 \, bar$ applying the resulting pressure to the drive line to control the VAD.}
\label{VAD}
\end{figure}

\paragraph{Heart Pump}
The simulator in-flow conditions at the heart node were realised by a pneumatically driven medical VAD (Ventricular Assist Device) diaphragm pump (Medos Stolberg, Germany) with maximum stroke volume of 80 ml, which provides a pulsatile fluid flow through the vascular system in a manner analogous to the left ventricle of the heart. The diaphragm pump is a medical device generally used as a external mechanical circulatory support system e.g. as bridge for heart transplant patients and therefore is suitable to create a pulsatile and heart-like pumping behaviour \citep{Thuaudet.2000}.
The diaphragm pump contains two heart valves and is controlled by a 3/2-way proportional valve (Series 614, Sentronic), which outputs the pressure for the drive line (see figure \ref{VAD}). The proportional valve applies the resulting pressure of a defined pressure curve by mixing an applied relative underpressure of $ p_u = 0,4 - 0,7 \, bar$ and overpressure of $p_o = 1 \, bar$. The vacuum pressure is generated by a pressure-controlled vacuum pump and stored in a recipient of 40 litre, to reduce long term drift during systole and realise long simulation times with stable pressure conditions.

During diastole the air side of the diaphragm pump is subjected to vacuum pressure reducing the air chamber volume, thus the membrane moves toward the air side and the ventricle is filled. The fluid is transported into the system by applying overpressure to push the medium out of the VAD through the arterial outlet.

\begin{figure} [H]
\centering
\includegraphics[width=\linewidth]{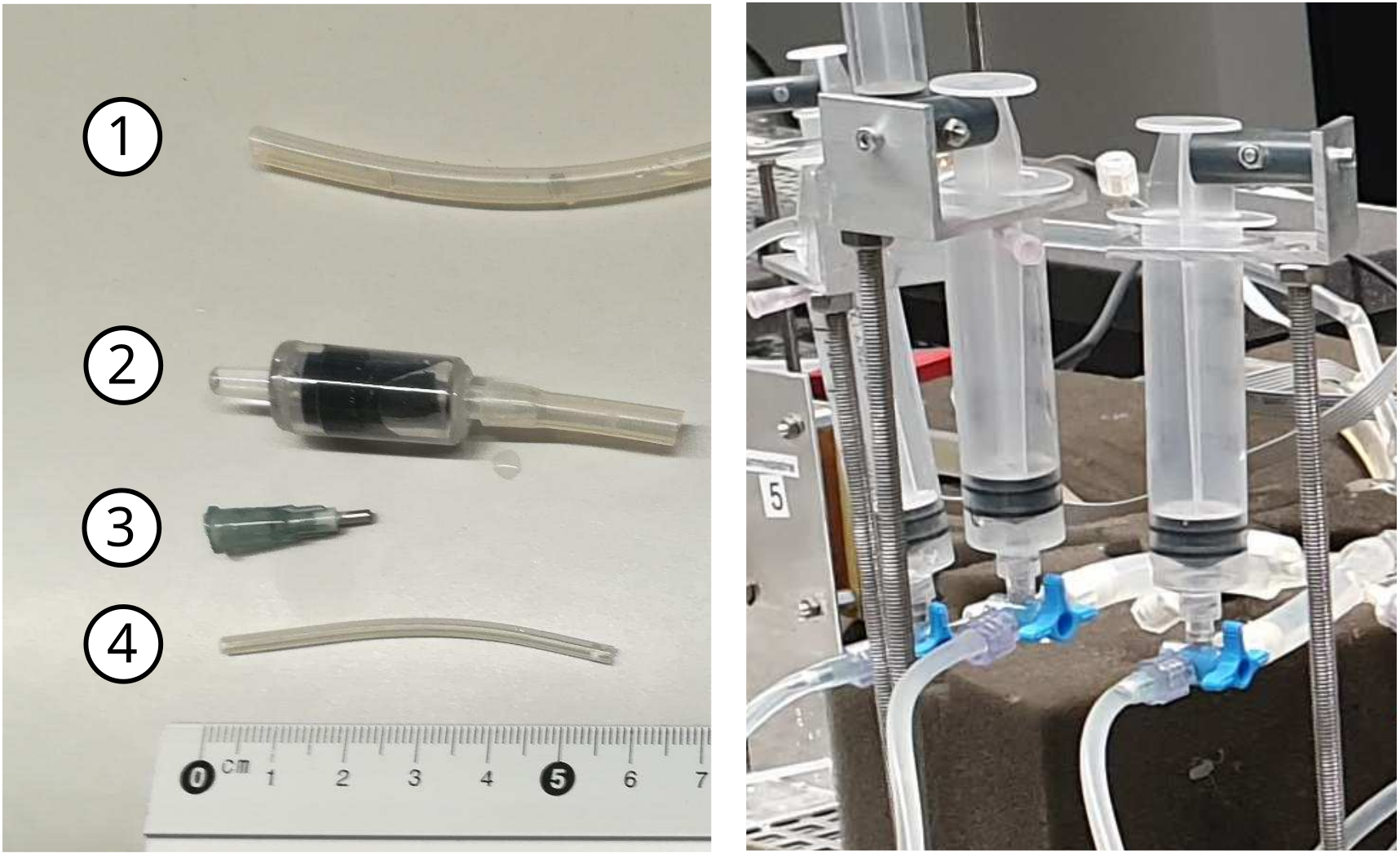}
\vspace{-3mm}
\caption{Peripheral resistance elements (left), including (1) outer tube, (2) valve, (3) cannula  and (4) small tube and compliance element (right) containing 3-way cock and a compliance syringe.}
\label{PeripheralElements}
\end{figure}

\paragraph{Peripheral Resistance and Compliance}
The physiological flow resistance of human arterioles and capillaries is modelled by additional small tubes inserted into the peripheral arteries, which open into the venous system (see figure \ref{PeripheralElements}). The peripheral resistance consists of the outer tube, a cannula with a small inner tube and a check valve. The length of the inner tubes was adjusted according to the physiological flow resistance of the arterial branch. Capillary flow resistance values were reproducibly generated downstream of each vessel end, the relative group values are found in table \ref{peripheralGroups}. Analogous to the venous valves in the human body, which prevent back flow in case of venous overpressure, for example, at the transition of the flow wave into the venous system, check valves were integrated to prevent back flow of fluid from the venous to the arterial system. The peripheral viscous flow resistance is defined as
\begin{align} \label{eqn:R_p}
R_p  = \dfrac{\Delta p}{q},
\end{align}
where $\Delta p$ is the pressure difference and $q$ represents the volume flow.\\
The peripheral resistances of the boundary nodes were measured by the definition of regional groups like legs, arms, organs, head, etc. Table \ref{peripheralGroups} shows the results in relation to the total peripheral resistance of the arterial system $R_p = (1,94 \pm 0,02) \cdot 10^8\, Pa\cdot s / m^{3}$ at the given viscosity (for detailed measurement description see appendix peripheral resistance measurement).

\begin{table}
\caption{\label{peripheralGroups}Measured peripheral resistance for each group in relation to the total peripheral resistance $R_p = (1,94 \pm 0,02) \cdot 10^8\, Pa\cdot s / m^{3}$ of the arterial system. }
\begin{tabular}{llr}
\hline
Group & Corresponding $R_p$ elements &$R_p^{-1} / R_{p_{tot}}^{-1}$ (\%)\\ 
% & & (\%)\\
\hline 
Head & $R_{p_{21}} , R_{p_{19}} , R_{p_{95}} , R_{p_{93}} $&17,52 \\
Coronar Art.& $R_{p_{2}} , R_{p_{3}}$& 5,57 \\
Arm dextra &$R_{p_{12}} , R_{p_{13}} $ &14,94 \\
Arm sinistra &$R_{p_{87}} , R_{p_{88}} $& 10,27 \\
Organs &$R_{p_{33}} , R_{p_{71}} , R_{p_{76}} , R_{p_{79}} , R_{p_{80}} $& 23,12 \\
Femoralis&$R_{p_{52}} , R_{p_{53}} , R_{p_{57}} , R_{p_{58}} $& 9,51 \\
Leg dextra&$R_{p_{47}} , R_{p_{49}} , R_{p_{50}} , R_{p_{43}} $& 10,44 \\
Leg sinistra&$R_{p_{66}} , R_{p_{67}} , R_{p_{68}} , R_{p_{69}} $&  8,63\\
\hline
\end{tabular}

\end{table}

Compensation and adjustments of the compliance were realised by syringes integrated vertically at the transition to the venous tube system (see figure \ref{PeripheralElements}). These are filled with a defined volume of air and thus create an artificial, additional distensibility of the respective vessels (all syringes were set to an air volume of $V_{px} = 2 \, ml$, except at the peripheral nodes: $V_{p3} = 3 \, ml, V_{p50} = 5 \, ml$ and $V_{p66} = 6 \, ml$ (see figure \ref{schematic}). The syringes can thus be considered as peripheral windkessel elements and have an impact on the total systems compliance.  The compliance is defined as the extensibility of a artery and can be calculated by
\begin{align}
C  = \dfrac{\Delta V}{\Delta p },
\end{align}
where $\Delta p$ is the change in pressure for a prescribed change in volume $ \Delta V$.  The total systems compliance $C = (0,32 \, \pm \, 0,01) \, ml/mmHg$ was measured by adding a defined volume to the arterial system using a syringe connected via a luer-lock connector (for details see appendix compliance measurement figure \ref{compliance}). 

\paragraph{Pressure and Flow Sensors}
The pressure and flow was measured as a function of time, i.e. $p(t)$ and $q(t)$ at different locations in the system. Therefore, 17 pressure sensors (AMS 5812, 0050-D-I, inverted, fluid resistant) and a clamp-on medical grade ultrasonic-based flow sensor (SONOFLOW CO.55/060) were used to measure the flow velocity and the pressure in the system at predefined locations (see figure \ref{schematic} and \ref{macsim}, label F1 and 1-17). Specific locations of the pressure and the flow sensors are shown in the schematic in figure \ref{schematic}. Prior measurement all sensors were calibrated, detailed measurement setup and calibration results are given in figure \ref{calibrationpressure} in the appendix.

\subsection{Measurement Setup and Procedure}
\begin{figure} [H]
\centering
\includegraphics[width=\linewidth]{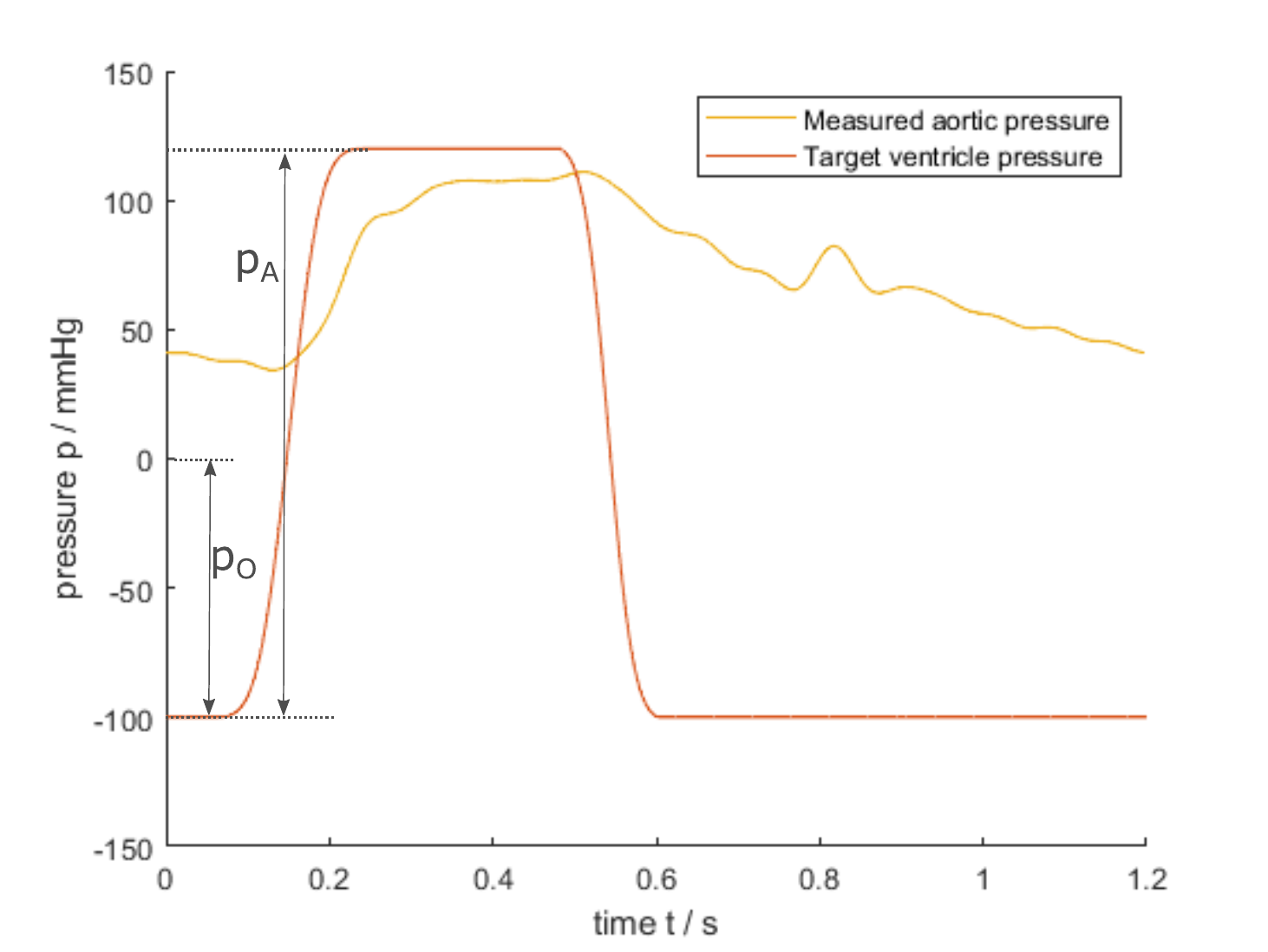}
\caption{Trapezoid VAD driving pressure (orange) was set between -100 mmHg and 120 mmHg, and smoothed by a gaussian windows of length 0,1 using matlab function smoothdata. Resulting aortic pressure, $p_{in}(t)$, at the root node of the vascular system (yellow).}
\label{inputcurve}
\end{figure}

For each measurement scenario the pressure and flow was measured at 17 + 1 predefined locations respectively (see figure \ref{schematic}) . The input heart curve was chosen to be a trapezoidal curve (see figure \ref{inputcurve}), which was parametrised by an amplitude, offset, heart frequency and start and end of the ascending/descending slope (see equation \ref{trapezfunction}). All measurements were acquired with a heart rate of $HR = 50 \, bpm$ and a maximum pressure amplitude of $p_A = 220 \, mmHg$ with an negative offset of $p_O = -100 \, mmHg$. 
The trapezoidal curve was generated on a normalized time scale $\tilde{t} = t/T$, where $T$ is the temporal period for the heart rate. 

\begin{equation} \label{trapezfunction}
p_{in}(\tilde{t}) = 
 \begin{cases} 
      p_{O} & 0 \leq \tilde{t} \leq \tilde{t}_{a,1} \\
      p_{O} + \frac{\tilde{t}-\tilde{t}_{a,1}}{\tilde{t}_{a,2}-\tilde{t}_{a,1}} p_{A} & \tilde{t}_{a,1} \leq  \tilde{t} \leq \tilde{t}_{a,2} \\
      p_{O} + p_{A} & \tilde{t}_{a,2} \leq  \tilde{t} \leq \tilde{t}_{d,1} \\
      p_{O} + p_{A} - \frac{\tilde{t}-\tilde{t}_{d,1}}{\tilde{t}_{d,2}-\tilde{t}_{d,1}} p_{A} &  \tilde{t}_{d,1} \leq  \tilde{t} \leq \tilde{t}_{d,2} \\
      p_{O} &  \tilde{t}_{d,2} \leq  \tilde{t} \leq 1
   \end{cases}  
\end{equation}

A linear raise was created between $\tilde{t}_{a,1} = 0,1$ and $\tilde{t}_{a,2} = 0,15$ followed by a plateau and a descent between $\tilde{t}_{d,1}=0,45$ and $\tilde{t}_{d,2}=0,5$. The resulting curve was smoothed by Matlabs \textit{smoothdata} function with a window length of 0,1 and rescaled along the time axis according to the applied heart rate (see figure \ref{inputcurve}).

The measurements were performed over a period of 60 seconds to guaranty steady state conditions and were acquired using a 16-bit data acquisition PCI-card (National Instruments, Austin, TX, USA) at sampling frequency of 1000 Hz per channel. The data acquisition software was entirely written in Matlab. The measurement data and meta-information was stored into a MySQL database for futher analysis.

\subsection{Measurement Scenarios}
\begin{figure} [H]
\centering
\includegraphics{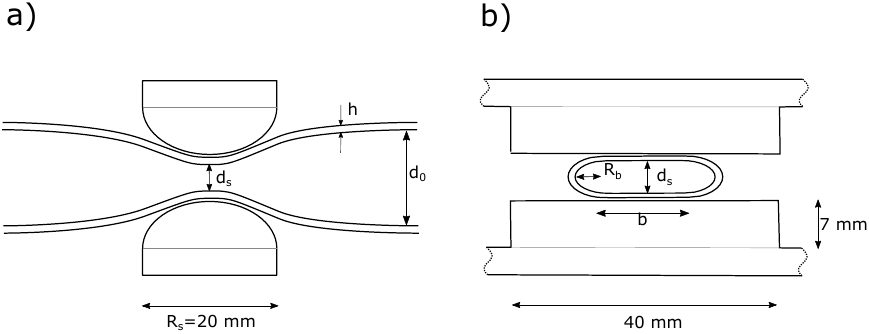} 
\caption{a) Axial cross-section of the 3-D printed parallel clamp to generate stenosis and the reduced vessel diameter.  Cross-section b) shows the vessels geometry in the stenosis region.}
\label{stenosisclamp}
\end{figure}

The influence of stenoses on the pressure and flow in the cardiovascular system was investigated by simulating different measurement scenarios under healthy and pathological conditions. The healthy state serves as the reference without artificial stenoses.
In the pathological setup an artificial stenosis in art. femoralis 20 cm downstream the knee (see figure \ref{schematic}) was chosen. The artery was squeezed reducing one axial dimension to a fraction between 3,3 \% and 25 \%, to obtain different degrees of stenosis (see table \ref{scenarios}). This setting does not directly correspond to the clinical situation, where the stenosis cross-section is circular. In this study, the percent reduction of the artery is defined through the area change and the change in the diameter. The shape of the stenosed artery in squeezed form (see figure \ref{stenosisclamp}) is described by a rectangle with two attached half circles \citep{Bernhard.2006}, then the cross-sectional area can be written as $A_2 = bd_s+(d_s/2)^2\pi$ where $b$ is the width of the rectangle and $d_s$ the squeezed inner diameter as seen in figure \ref{stenosisclamp}. For negligible bending resistance in a thin walled tube, the circumference remains unchanged when squeezing the tube, in this case one can express the ratio $A_2/A_1$ as a function of the ratio $\delta = d_s/d_0$, where $A_1=(d_0/2)^2 \pi$ is the cross-sectional area of the unsqueezed artery and $d_0$ is the initial inner diameter:
\begin{align}
\dfrac{A_2}{A_1} = 2\delta -\delta^2,
\end{align}
for $\delta \in [0,1]$.

\begin{table}[H]
\caption{\label{scenarios}Definition of the measurement scenarios of stenosis at art. femoralis dextra with different area and diameter reduction. $\delta$ refers to the reduction of the diameter and $A_2 / A_1$ is the fraction of the reduction of the vessels area. }
\vspace{0.5cm}
\begin{tabular}{ccc}
\hline
No.   & $\delta$  & $A_2 / A_1$ \\
\hline 
I    & 100 \% & 100 \% \\
II  &   25 \%  & 37,5 \% \\
III  &   12,5 \% & 23,4 \% \\
IV &  3,3 \% & 6,56 \% \\
\hline
\end{tabular}
\end{table}

All stenoses were established using a 3D-printed clamp (see figure \ref{macsim} for the printed object and figure \ref{stenosisclamp} for cross-section).

%------------------------------------------------

\section{Results}

The resulting data set is structured into four mat files, one per scenario. Each file contains 18 pressure signals and one flow signal, in total the data set contains 76 signals. The dataset and a detailed description is available at \citep{Bernhard.2022}. The following subsections describe the properties and results.

\subsection{Pressure waves along the arterial network}
\begin{figure} [H]
\centering
\includegraphics[width=\linewidth]{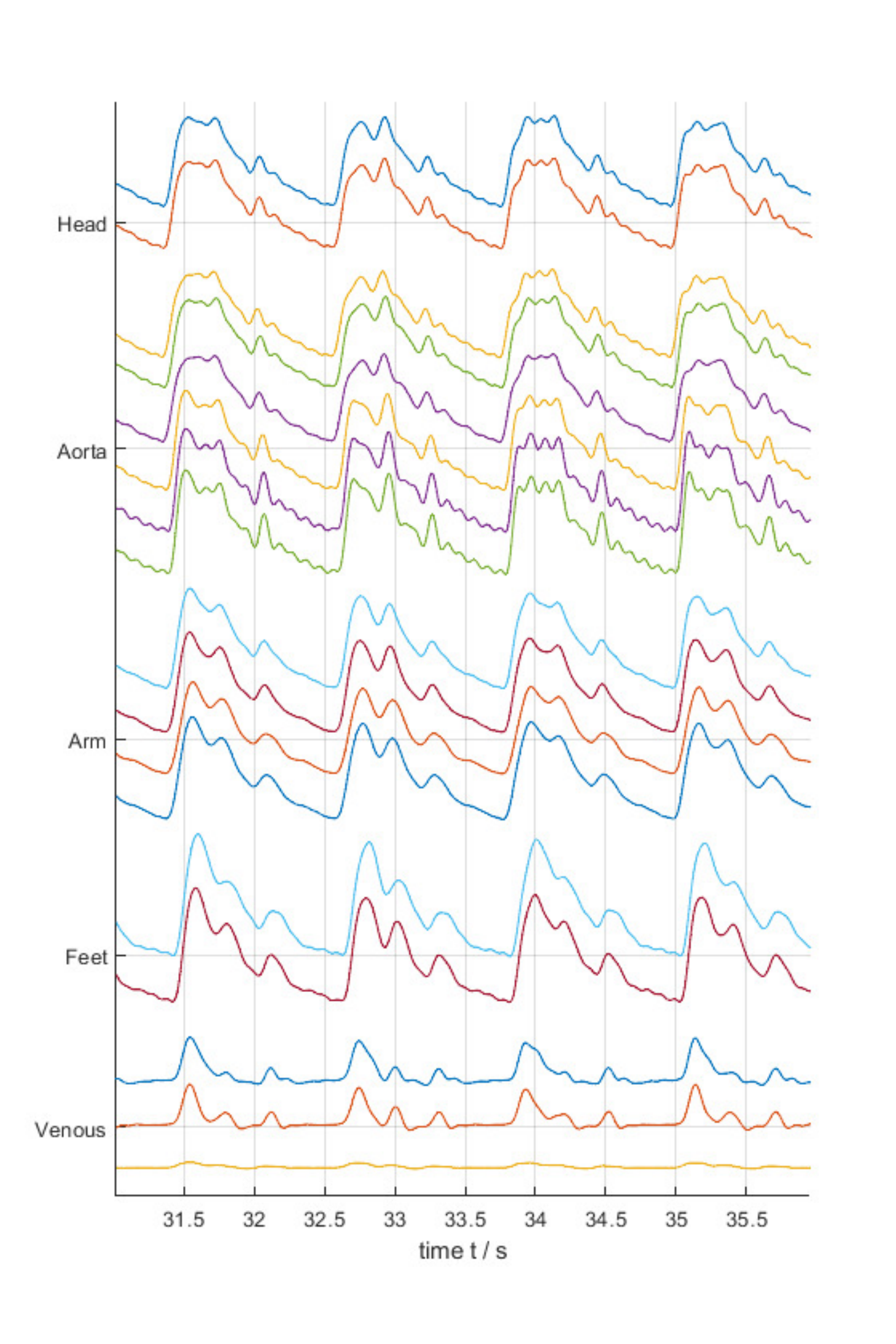}
\caption{Entire set of pressure waves along the arterial network under healthy conditions (scenario I), including venous return path.}
\label{allpressurecurves}
\end{figure}

Figure \ref{allpressurecurves} shows the entire set of pressure curves along the arterial system under healthy conditions. Due to wave reflections of discontinuities the pressure waves clearly change their shape while propagating through the arterial system. As expected a short time delay between aortic and peripheral waves is observed (transit-time), which manifest, according to the wave velocity in the arterial network. The pressure amplitude increases in the peripheral vessels, which is in agreement with the pulse wave amplification observed in in-vivo measurements.

\subsection{Scenario I - Healthy Conditions}
\begin{figure} [H]
\centering
\includegraphics[width=\linewidth]{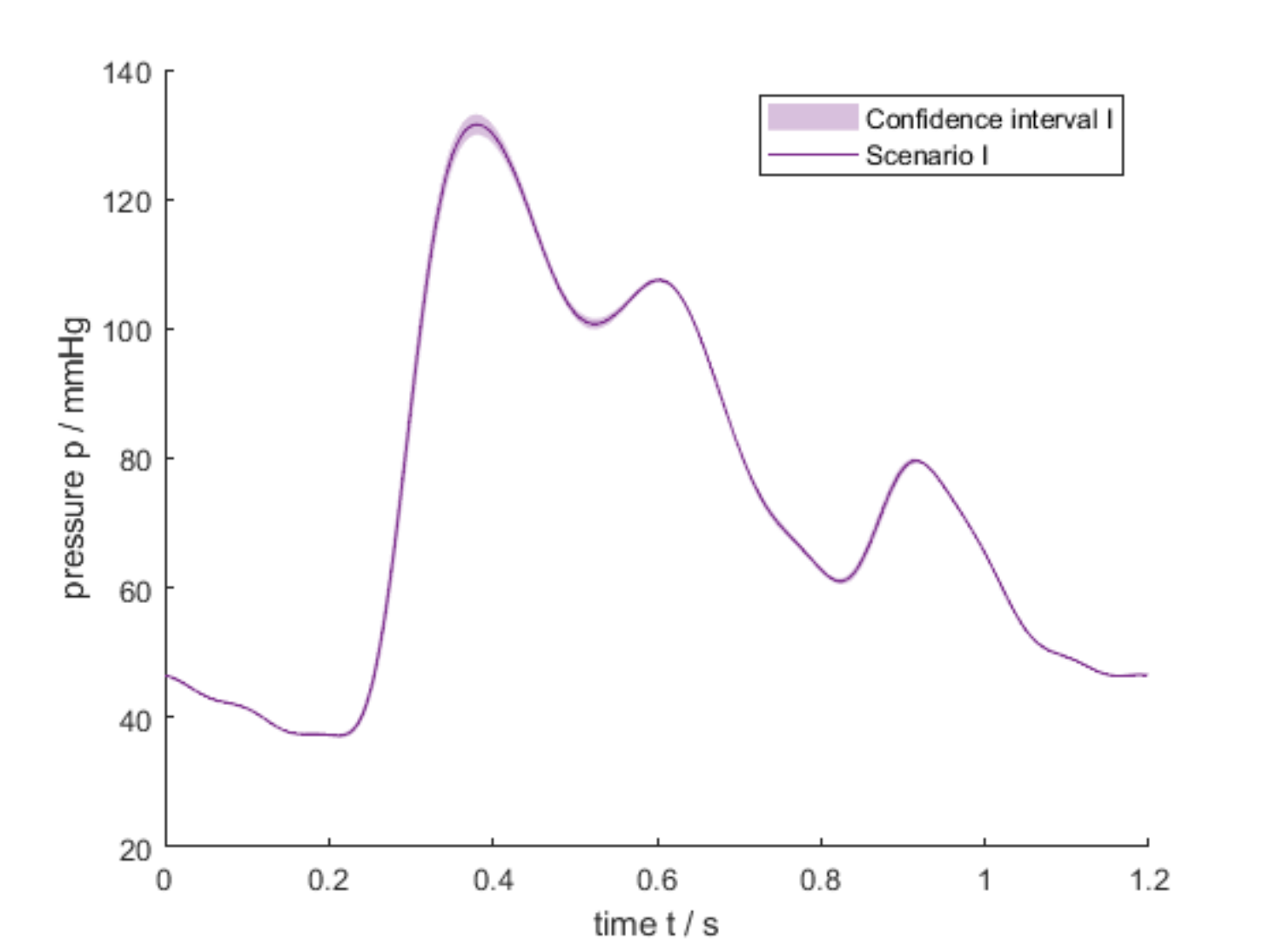}
\caption{Pressure wave at art. tibialis dextra under healthy conditions (scenario I). The confidence interval was computed over five periods.}
\label{pressurehealthy}
\end{figure}

In figure \ref{pressurehealthy} the pressure wave at art. tibialis dextra under normal physiological conditions is shown. The result is similar to in-silico simulations and literature in terms of waveshape and specific wave features like the dicrotic notch and peripheral steeping.

After the systolic rise to the peak pressure of $\hat{p}_{I} = 132 \, mmHg$, the blood pressure drops until the aortic valve closes, resulting into a dicrotic notch in the decaying pressure wave. This notch (incisure) is also found in human pressure waves. Subsequently, the pressure falls down to diastolic level of about $\check{p}_{I} = 37,2 \, mmHg$, which is much lower than it would be physiologically.\\

The figures \ref{pressurehealthy} to \ref{flowpat} contain confidence intervals calculated by the standard deviation of coherent averages, i.e. five averaging windows of the size of eight periods were used. The intervals represent the point-wise standard deviation and are used to show the temporal variation within the pressure waves. The confidence interval along the pressure waves is small, but increases at the systolic peak values and the discrotic notch. The mean value of the standard deviation of the systolic/diastolic peak values for pressure and flow for each scenario are given in table \ref{resultsPressFlow}.\\

\subsection{Scenarios II-VI - Pathological Conditions}
\begin{figure} [H]
\centering
\includegraphics[width=\linewidth]{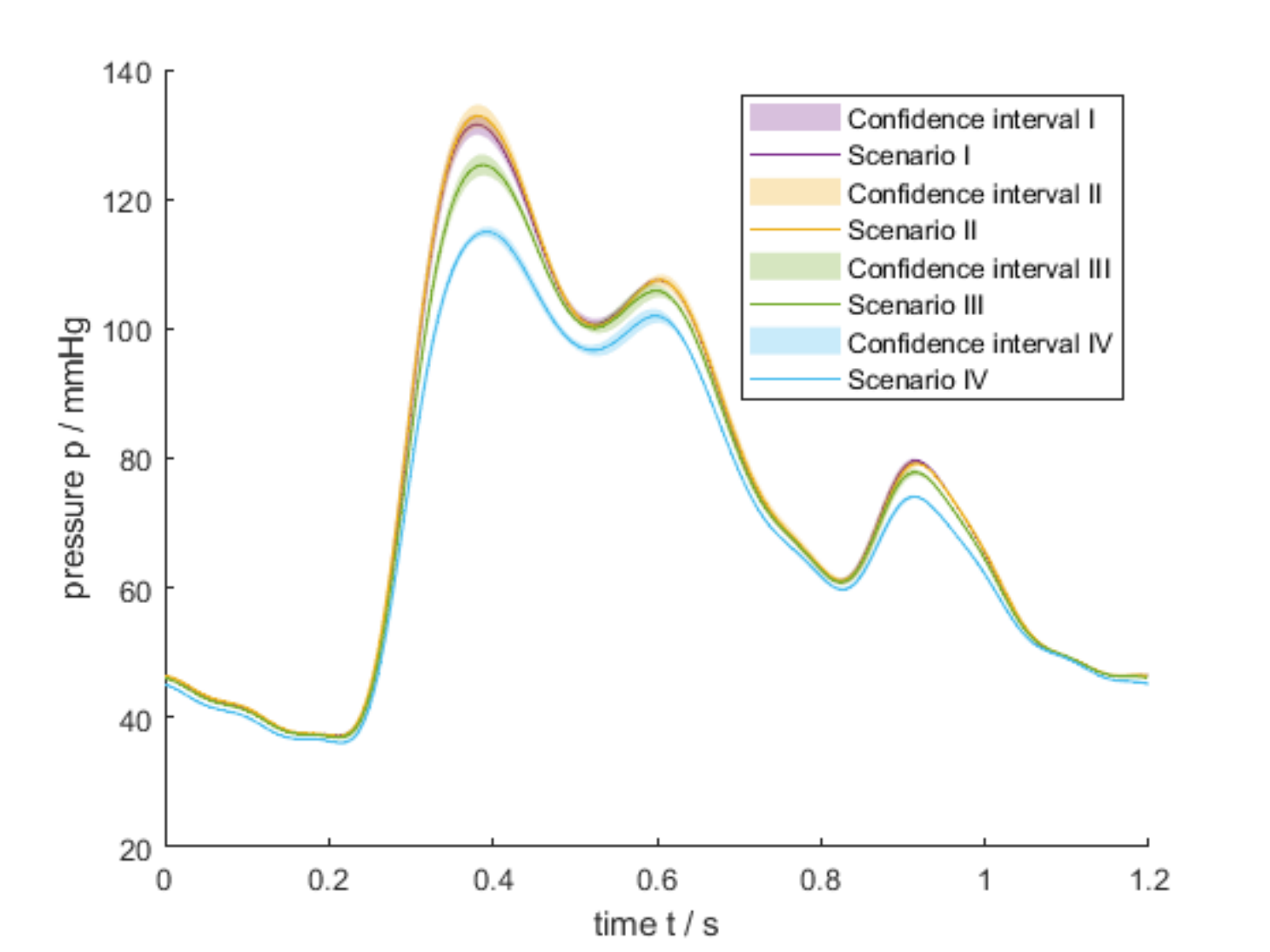}
\caption{Pressure waves for scenario I-IV at art. tibialis dextra.}
\label{pressurepat}
\end{figure}

\begin{figure} [H]
\centering
\includegraphics[width=\linewidth]{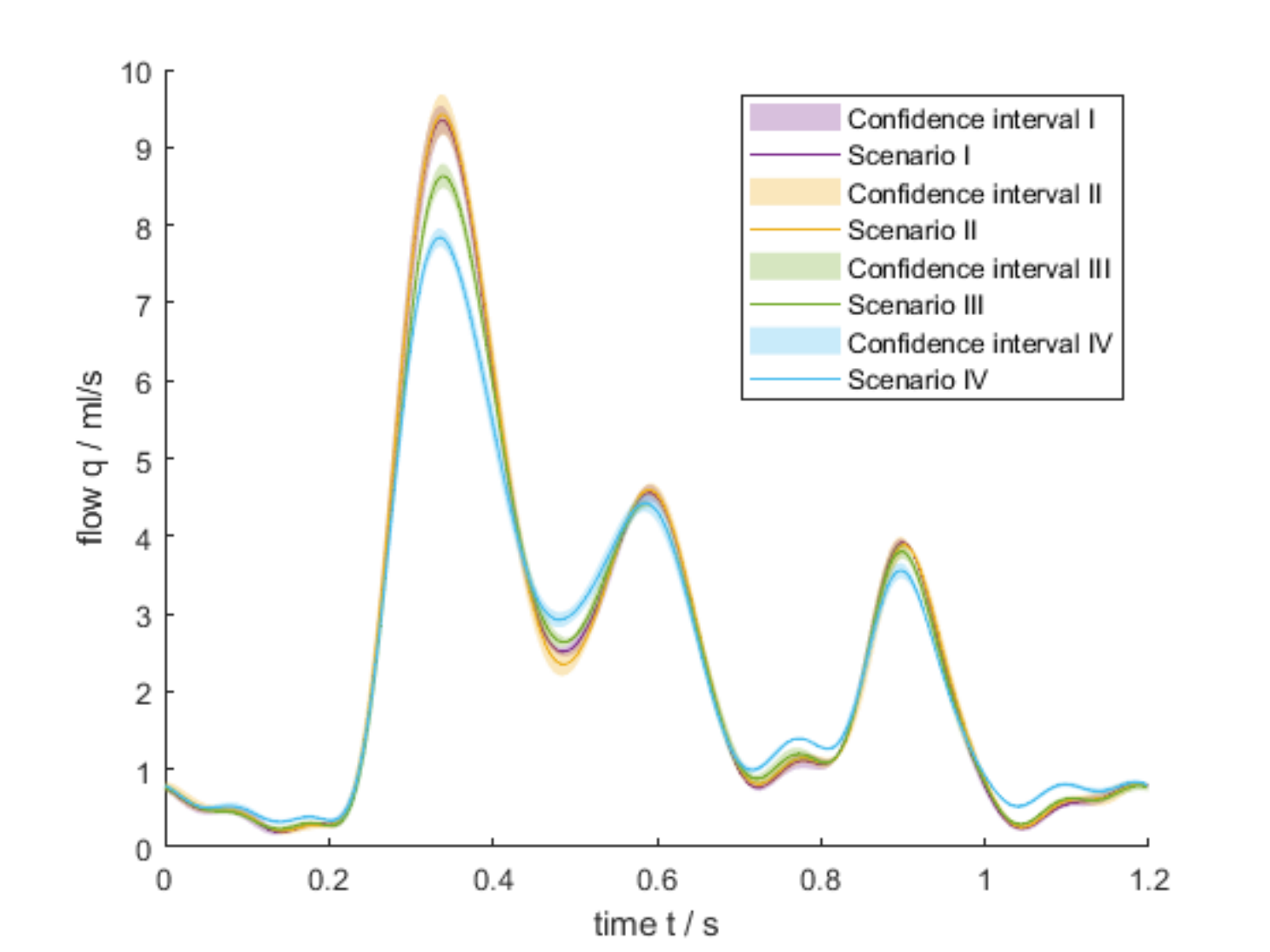}
\caption{Flow waves of scenario I-IV at art. profunda femoris dextra.}
\label{flowpat}
\end{figure}

The pathological conditions II-VI are based on a stenosis in the art. femoralis dextra with different stenosis degree (see table \ref{scenarios}), corresponding the measurement result is given in figure \ref{pressurepat} and \ref{flowpat}.

The pathological scenario II contains a stenosis in art. femoralis with a stenosis degree of $\delta_{II} =  25 \%$. Due to the low degree of the stenosis there is no significant difference in the characteristics of the pressure wave, as expected the stenosis has low effect on the blood pressure: The pressure increases to a systolic peak value of $\hat{p}_{II} = 133 \, mmHg$ and diastolic peak value of $\check{p}_{II} = 37,4 \, mmHg$ is observed.

The pathological scenario III contains a stenosis with a higher degree of $\delta_{III} =  12,5 \%$, which causes a decrease of the pressure peak values of the pulse wave at the art. tibialis dextra (see figure \ref{pressurepat}). The systolic pressure peak decreases by 6 mmHg to an amplitude of $\hat{p}_{III} = 126 \, mmHg$, while the diastolic pressure remains constant at $\check{p}_{III} = 37 \, mmHg$. Compared to the healthy setup, the shape of the pulse waves distal to the stenosis smoothes due to the reduction of the vessel’s effective diameter by the constriction. As expected the scenario IV has the lowest systolic pressure of all scenarios. In comparison to the reference scenario I the systolic pressure  significantly decreases by 16 mmHg to a peak value of $\hat{p}_{IV} = 115,8 \, mmHg$. 

The mean pressure values for each scenario are given in table \ref{resultsPressFlow}. With increasing stenosis degree the mean pressure $\overline{p}$ decreases, but not that strong as the peak values $\hat{p}$. The difference of the mean pressure between scenario I and IV is only 4,3 mmHg, which can explained by the fact that although the systolic pressure decreases, the diastolic pressure remains at the same level for all scenarios.

In figure \ref{flowpat} the flow waves at the art. profunda femoris dextra are shown for all measurement scenarios. The peak values of the flow velocity for the healthy state (scenario I) is $\hat{q}_{I} = 9,4 \, ml/s$, and for all pathological conditions reduced as expected $\hat{q}_{II} = 9,3 \, ml/s$, $\hat{q}_{III} = 8,6 \, ml/s$ and $\hat{q}_{IV} = 7,8 \, ml/s$. Consequently, the flow velocity within the diseased vessel decreases with an increasing degree of the stenosis. The mean flow values for each scenario are given in table \ref{resultsPressFlow}. In contrast to the peak values the mean flow remains almost constant.

\begin{table*}[btp]
\caption{\label{resultsPressFlow}Results of the measurement scenarios regarding pressure and flow amplitudes. $\hat{p}$ refers to the systolic and $\check{p}$ to the diastolic pressure, while $\hat{q}$ refers to the peak value of the flow wave.  $\overline{p}$ and $\overline{q}$ are the mean value of pressure and flow, while $\overline{STD_p}$ and $\overline{STD_q}$ are their mean standard deviations, respectively.}
\centering
\vspace{0.5cm}
\begin{tabular}{cccccccc}
\hline
No. & $\hat{p}$ (mmHg) & $\check{p}$ (mmHg) & $\overline{p}$ (mmHg) & $\overline{STD_p}$ (mmHg) & $\hat{q}$ (ml/s)& $\overline{q}$ (ml/s) & $\overline{STD_q}$ (ml/s)\\
\hline 
I    	& 132,0  	&37,2	& 73,7	& 0,7 & 9,4 &2,4	&0,1\\
II  	& 133,0  	&37,4	& 73,9 	& 0,8 & 9,3 &2,4	&0,1\\
III  	& 126,0 	&37,0	& 72,4	& 0,7 & 8,6 &2,3	&0,1\\
IV 		& 115,8		&36,2	& 69,4	& 0,6 & 7,8 &2,3	&0,1\\
\hline
\end{tabular}
\end{table*}

\begin{figure} [H]
\centering
\includegraphics[width=\linewidth]{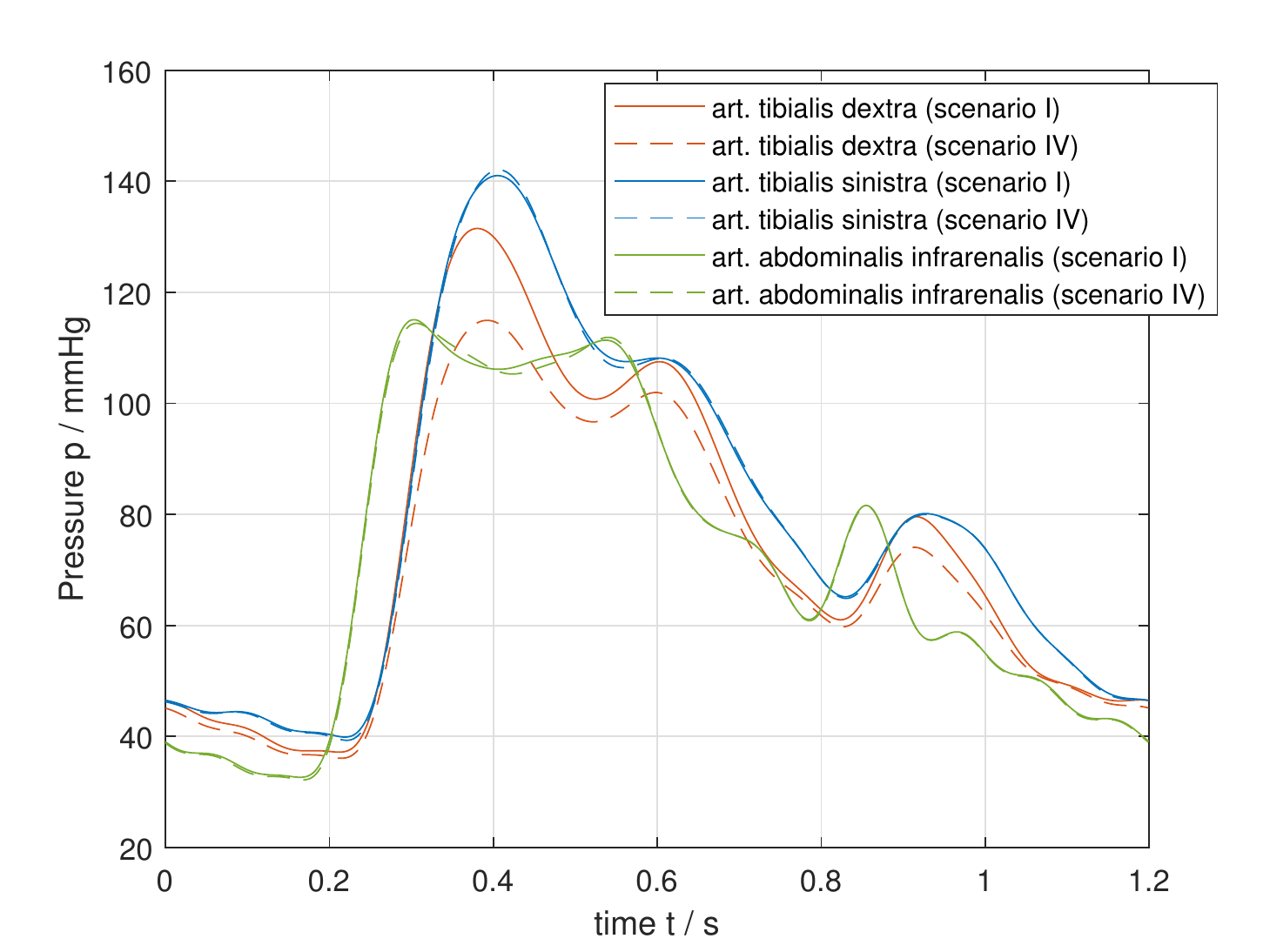}
\caption{Impact of the stenosis in the art. femoralis dextra on the pressure waves in the aorta and the feet.}
\label{stenosisDextraVsSinistra}
\end{figure}

Finally, figure \ref{stenosisDextraVsSinistra}, shows the influence of the stenosis on different adjacent arteries like art. abdominalis infrarenalis and art. tibialis dextra and sinistra. In scenario I without stenosis $\delta_{I} = 100 \%$, while in scenario IV $\delta_{IV} = 3,3 \%$. The pressure wave measured in the right foot decreases, while the pressure measured in the aorta and in the left foot remain visually unchanged.

%------------------------------------------------

\section{Discussion}
The purpose of this study was the development and validation of a patient-specific in-vitro hardware simulator to generate parametric data sets under healthy and pathological conditions for computational model validation. 

In the past years, different hardware simulators were developed to investigate various theses, thus the three dimensional arterial structure differs in  complexity, in the type of heart pump, in the number of sensors and in properties like compliance and peripheral resistances.
In \citep{Gehron.2019}, the simulator drive consists of two pneumatically driven VADs representing the right and the left ventricle. In \citep{Matthys.2007}, a harvard pulsatile pump is used. Furthermore, both hardware simulators \citep{Gehron.2019, Matthys.2007} contain  detailed arterial network covering 37 major arteries of the human body. Whereas hardware simulators with the aim to test and validate VADs, include an arterial network of low complexity and a simple functional drive.

The simulator in this work is pneumatically driven by one VAD to provide a pulsatile fluid flow through the vascular system. The arterial network contains a detailed arterial network with adjustable elements regarding heart rate, systolic/diastolic pressure, compliance and peripheral resistances. Furthermore, the simulator provides 17 pressure sensors at different locations and one flow sensor, which enables a detailed evaluation of the wave propagation. Due to material properties the total arterial compliance of the simulator is $C = (0,32 \pm 0,01) \, ml/mmHg$ and therefore lower than in-vivo.
%($C_{human} = 1 \, ml/mmHg$)%
The total peripheral resistance is $R_p = (1,94 \, \pm \, 0,02) \cdot 10^8\, Pa\cdot s / m^{3}$.
%($R_{p_{human}} = 11,3–17,5 \, mmHg \cdot min/l$)
The low compliance evokes that the stiffness of arteries is higher than in-vivo, which indicates an atherosclerotic, high-blood-pressure patient. In contrast the arterial compliance in \citep{Gehron.2019} was adjusted to $1,0 \, ml/mmHg$ and in \citep{Matthys.2007} no peripheral compliances are included. 

 As shown in the results section, the pressure waves within the simulator contain similar properties as in-vivo measurements. The waveshape and specific wave features like the discrotic notch, peripheral steeping and translational pressure drop are observed. Furthermore, due to wave reflection at discontinuities and compliance variation of the vessels the shape of the pressure changes while propagating through the system.
  
Moreover, the influence of a stenosis and its degree on the pulse wave in the circulatory system was investigated. As expected, the results imply that for higher degree of stenosis the pressure after the diseased vessel decreases. The flow measurements show similar results (see figure \ref{flowpat}): for higher degree of stenosis the flow decreases, as expected. These results are in good conformity to results provided by other hardware simulators \citep{Hacham.2019, Jin.2021}.
Furthermore, the influence of stenosis on different arteries were examined. Figure \ref{stenosisDextraVsSinistra} shows that the stenosis at art. femoralis dextra has only an significant impact on the diseased vessel in the right leg, where the pressure decreases. Thereby, the pressure wave in the aorta and the left leg remain visually unchanged. The measurements under physiologic and pathological conditions confirm the validity of the in-vitro hardware simulator.

However, certain limitations concerning the results of this study could be addressed in future research. A first limitation concerns the low compliance of the arterial system as well as the peripheral resistances, which are too high. The compliance could be adjusted by a higher volume of the syringes within the system. To get more physiological peripheral resistances the impact of the different resistance elements like inner tubes and valves could be adjusted.
A further potential limitation are the measured pressure waves, which are affected by noise in a broad frequency range. The reason for this are the vibrations of the systems components due to pumping process of the VAD. This could be fixed in future research by a more efficient embedding of the tube system with the PU-foam blocks. Moreover, the diastolic pressure of about 40 mmHg is too low in comparison with in-vivo measurements. The reason for this may be the low compliance and the high peripheral resistances within the system.

In conclusion, the present study has provided measurement data to the community, which hopefully provides support for the validation of computational models. In addition, the improvement of the pathological understanding will enable interpretation in a clinical setting through validation of computational models.

In terms of future research, it would be useful to extend the current findings by generating a data set with the hardware simulator developed, that can be used to develop and test algorithms for stenosis detection and localization on physical in-vitro data.

\section{Conclusion}
Within this study a in-vitro cardiovascular hardware simulator was developed and validated to gain a deeper understanding of blood pressure and flow under healthy and pathological conditions.

Physiological flow conditions are adjustable in a wide range by changing parameters like heart rate, systolic/diastolic pressure, compliance and peripheral resistances. The pressure and flow waves show similar wave form compared to in-vivo measurements. Moreover, the pressure and flow waves show the expected behaviour, in case of a stenosis of different location and degree.

This work provides measurement data containing healthy and pathological conditions like stenoses to the research community, to support the validation of computational models in near future.

\section*{Declaration of competing interest}

All authors declare that there is no conflict of interest in this work.

%----------------------------------------------------------------------------------------
%	REFERENCE LIST
%----------------------------------------------------------------------------------------

\bibliography{MacSimReferences.bib}

\newpage
\newpage
\section*{Appendix}

\subsection*{Calibration Measurements}
All pressure and flow sensors used in this study were calibrated to ensure valid measurement data.

\subsection*{Calibration Pressure Sensors}
\begin{figure}[H]
\centering
\includegraphics[width=\linewidth]{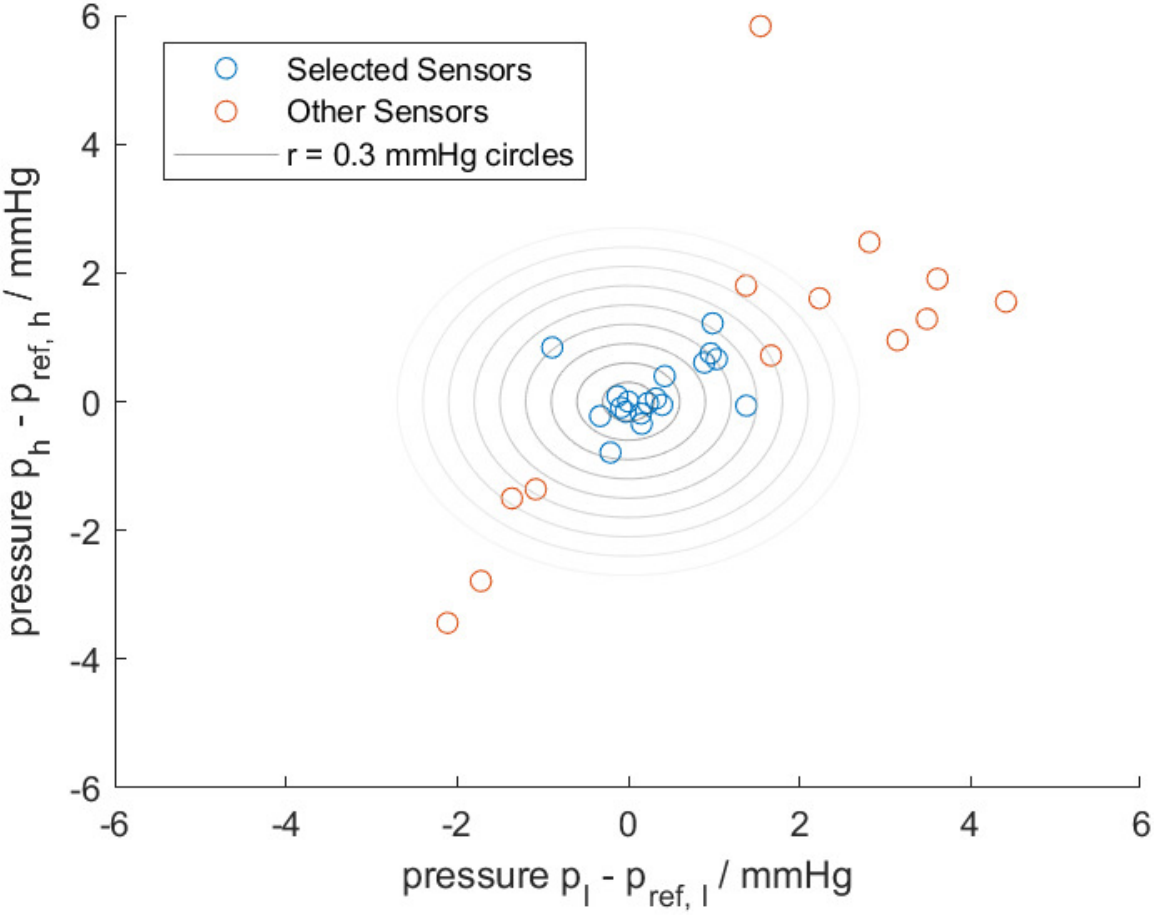}
\caption{Scatter plot of calibrated pressure sensors in comparison to the reference sensor $p_{ref}$.}
\label{calibrationpressure}
\end{figure}

The pressure sensors were calibrated through a two-point calibration measurement. Therefore, a bag filled with water was set to a defined hydrostatic pressure. This pressure corresponds to a water column of $p_h = 820\, mmH_2O = 61,8 \,mmHg$. Subsequently, the hydrostatic pressure was set to $p_l = 0 \,mmHg$ compared to the atmospheric pressure for the second point for the calibration measurement.
In each calibration measurement a reference sensor, $p_{ref}$, was present to compare the measurement values. The results of the calibration measurement for each sensor is shown in figure \ref{calibrationpressure}. All sensors used lie within a maximum deviation of $\pm 1,5 \, mmHg$.

\subsection*{Calibration Flow Sensor}
The flow sensor was calibrated through a two-point calibration measurement, where the volume difference between the steady state and running system was evaluated at location F1 (see figure \ref{schematic}). Volume integration was done by disconnection of reservoirs and determination of the fluid amount per time. Subsequently, the mean flow velocity and a correction factor of 4,8 were calculated.

\newpage
\subsection*{Compliance}
The system compliance was evaluated by measurement of the pressure changes resulting from consecutive fluid injections of $\Delta V = 10 \, ml$  into the closed arterial system (see table \ref{tab:compliance}).  

\begin{table}[H]
\caption{Pressure-volume measurements and compliance evaluation of the arterial system obtained by consecutively fluid injections of $\Delta V = 10 \, ml$.}
\begin{tabular}{cccc}
\hline
No. & $\Delta \overline{p}$ (mmHg) & $\Delta V$ (ml) & C (ml/mmHg)\\
\hline
1 & 32,7 & 10 & 0,3058 \\
2 & 30,7 & 10 & 0,3257 \\
3 & 30,3 & 10 & 0,3300 \\
4 & 30,3 & 10 & 0,3300 \\
\label{tab:compliance}
\end{tabular}
\end{table}

\begin{figure}[H]
\centering
\includegraphics[width=\linewidth]{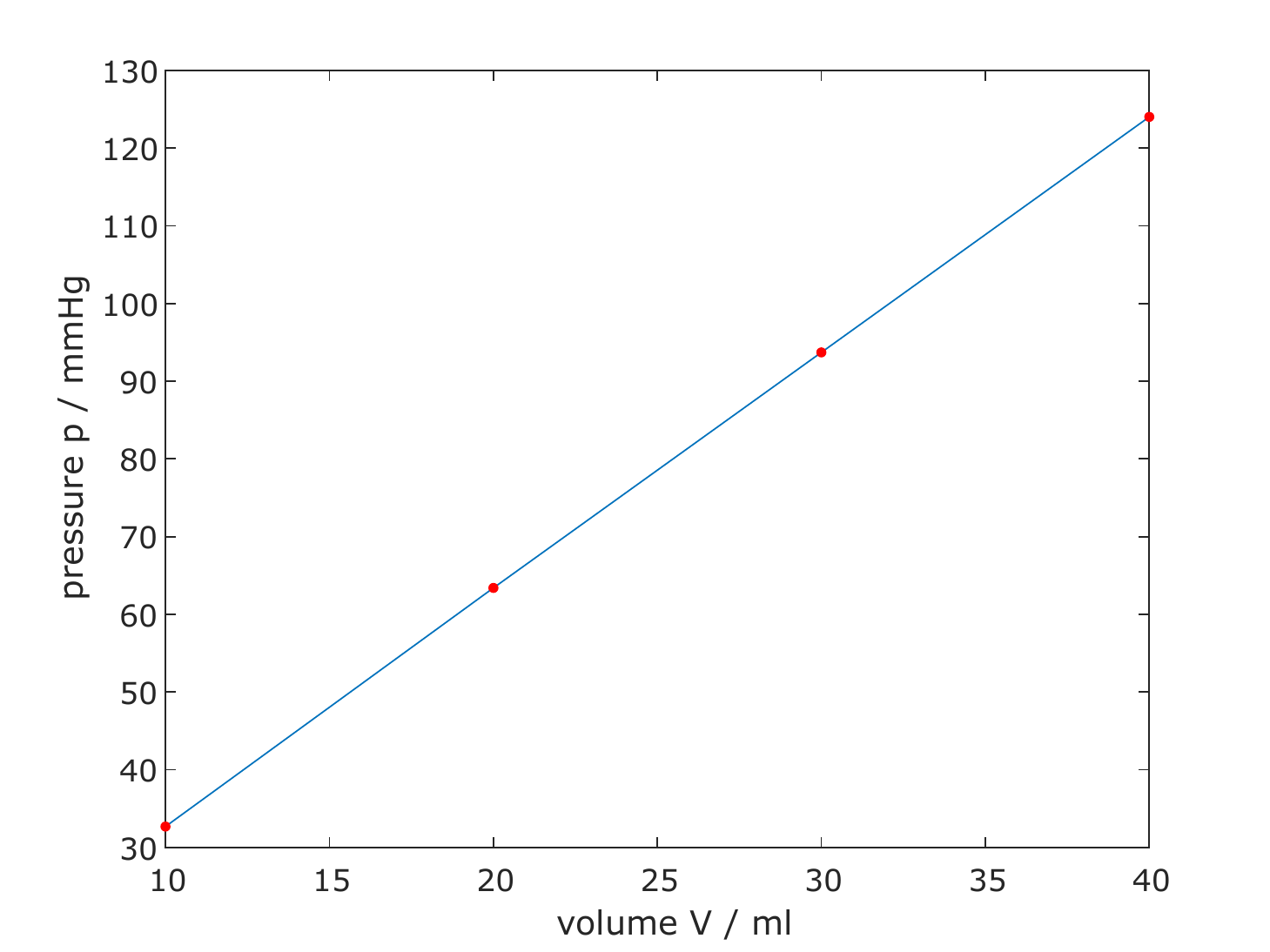}
\caption{Pressure-volume relation of the arterial network for injection of volume of $\Delta V = 10 \, ml$. The total arterial compliance refers to the slope of the curve.}
\label{compliance}
\end{figure}

The resulting pressure-volume relation is plotted figure \ref{compliance}, the linear slope implies proportional relationship in the measurement region as expected.  Consequently the total arterial compliance can be calculated by equation \ref{eqn:compliance} using the mean pressure difference $ \overline{p}$.
\begin{align} \label{eqn:compliance}
C  = \dfrac{\Delta V}{\Delta \overline{p}} = \dfrac{10 ml}{31 mmHg} = 0,32 \, ml/mmHg
\end{align}

\newpage
\subsection*{SISCA Model}
\begin{figure}[H]
\centering
\includegraphics[width=\linewidth]{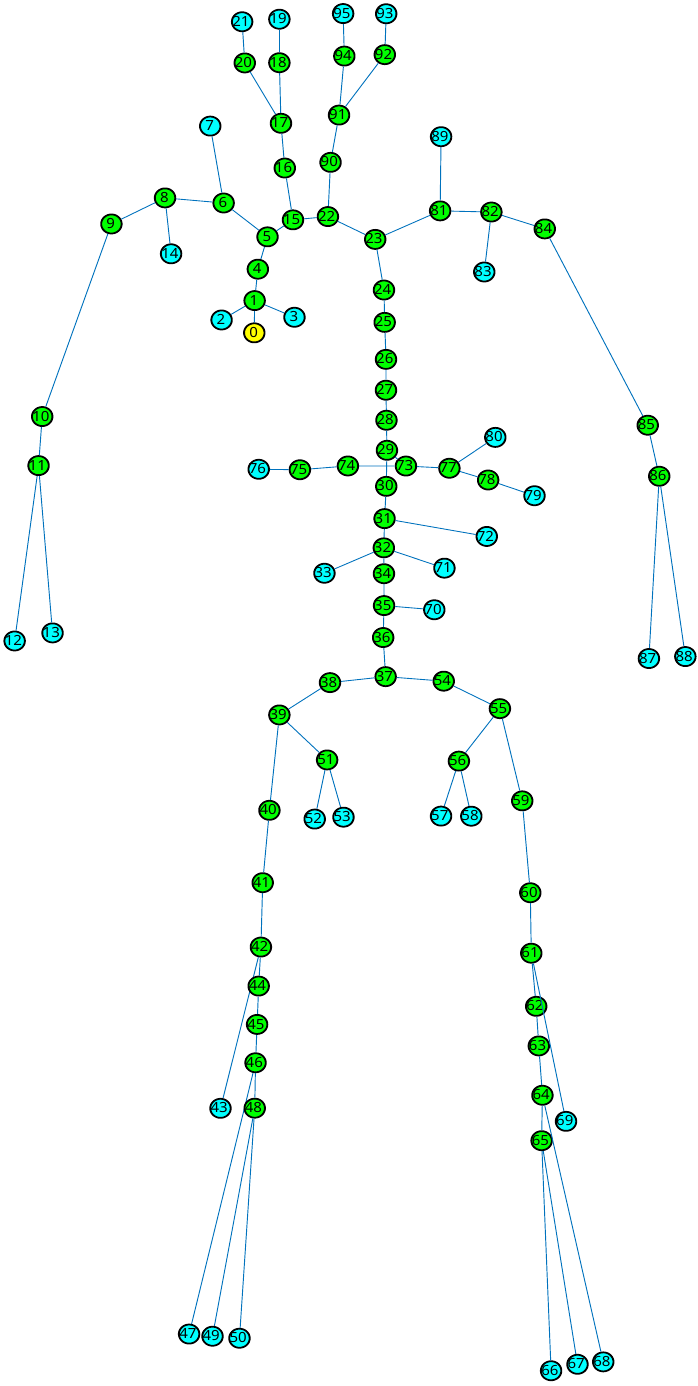}
\caption{SISCA network structure of the MACSim.}
\label{sisca}
\end{figure}

The node structure of the hardware simulator refers to a computational simulation model realised in the SISCA modelling environment \citep{Huttary.2017}. The node numbering of the arterial tree in SISCA is realised by a depth first search. tree The SISCA software and the simulation model (shown in figure \ref{sisca}) are available at  \citep{.06.04.2022}.

\newpage

\subsection*{Overview structural properties of the MACSim}
Table \ref{structuralMeasures} shows an overview of the properties length l, diameter d, wall thickness h and elasticity E of the corresponding SISCA node ID.

\begin{table}[H]
\caption{\label{structuralMeasures}Structural properties of the MACSim corresponding to SISCA node IDs}
\vspace{0.5cm}
\begin{tabular}{ccccc}
\hline
Node ID &   l (m) & h (m) & d (m)  & E (Pa) \\ 
\hline 

  		1 & 0.02 & 0.0015 & 0.025 	& 6700000 \\ 
        2 & 0.067 & 0.0005 & 0.004 	& 1650000 \\ 
        3 & 0.08 & 0.0005 & 0.004 	& 1650000 \\ 
        4 & 0.01 & 0.0015 & 0.025 	& 6700000 \\ 
        5 & 0.052 & 0.0015 & 0.025 	& 6700000 \\ 
        6 & 0.049 & 0.001 & 0.01 	& 6700000 \\ 
        7 & 0.14 & 0.0005 & 0.0015 	& 1650000 \\ 
        8 & 0.016 & 0.001 & 0.01 	& 6700000 \\ 
        9 & 0.053 & 0.001 & 0.01 	& 6700000 \\ 
        10 & 0.35 & 0.001 & 0.006 	& 1650000 \\ 
        11 & 0.021 & 0.001 & 0.004 	& 1650000 \\ 
        12 & 0.325 & 0.001 & 0.004 	& 1650000 \\ 
        13 & 0.345 & 0.0005 & 0.003 & 1650000 \\ 
        14 & 0.095 & 0.0004 & 0.0015 & 1650000 \\
        15 & 0.013 & 0.0015 & 0.025 	& 6700000 \\ 
        16 & 0.11 & 0.0005 & 0.0065 	& 6700000 \\ 
        17 & 0.045 & 0.0005 & 0.0065 	& 6700000 \\ 
        18 & 0.054 & 0.0005 & 0.004 	& 6700000 \\ 
        19 & 0.036 & 0.0005 & 0.004 	& 6700000 \\ 
        20 & 0.06 & 0.0005 & 0.006 		& 6700000 \\ 
        21 & 0.029 & 0.0005 & 0.006 	& 6700000 \\ 
        22 & 0.012 & 0.0015 & 0.028 	& 6700000 \\ 
        23 & 0.01 & 0.0015 & 0.028 		& 6700000 \\ 
        24 & 0.002 & 0.0015 & 0.028 	& 6700000 \\ 
        25 & 0.05 & 0.0015 & 0.025 		& 6700000 \\ 
        26 & 0.05 & 0.0015 & 0.021 		& 6700000 \\ 
        27 & 0.05 & 0.0015 & 0.02 		& 6700000 \\ 
        28 & 0.049 & 0.0015 & 0.019 	& 6700000 \\ 
        29 & 0.027 & 0.0015 & 0.019 	& 6700000 \\ 
        30 & 0.02 & 0.0015 & 0.018 		& 6700000 \\ 
        31 & 0.006 & 0.0015 & 0.017 	& 6700000 \\ 
        32 & 0.028 & 0.0015 & 0.016 	& 6700000 \\ 
        33 & 0.08 & 0.0005 & 0.004 		& 1650000 \\ 
        34 & 0.021 & 0.0015 & 0.016 	& 6700000 \\ 
        35 & 0.031 & 0.0015 & 0.015 	& 6700000 \\ 
        36 & 0.018 & 0.0015 & 0.015 	& 6700000 \\ 
        37 & 0.015 & 0.0015 & 0.014 	& 6700000 \\ 
        38 & 0.041 & 0.0001 & 0.01 		& 6700000 \\ 
        39 & 0.02 & 0.0001 & 0.01 		& 6700000 \\ 
        40 & 0.094 & 0.0001 & 0.01 		& 6700000 \\ 
        41 & 0.015 & 0.0001 & 0.01 		& 6700000 \\ 
        42 & 0.039 & 0.0001 & 0.008 	& 1650000 \\ 
                                 
\end{tabular}                     
\end{table}                     
                               
\begin{table}[H]                
\begin{tabular}{ccccc}          
\hline                           
Node ID &   l (m) & h (m) & d (m)  & E (Pa) \\ 
\hline                          
 43 & 0.28 & 0.0005 & 0.003 & 1650000 \\
        44 & 0.13 & 0.0005 & 0.008 		& 1650000 \\ 
        45 & 0.34 & 0.0005 & 0.006 		& 1650000 \\ 
        46 & 0.035 & 0.001 & 0.004 		& 1650000 \\ 
        47 & 0.425 & 0.0005 & 0.002 	& 1650000 \\ 
        48 & 0.049 & 0.001 & 0.004 		& 1650000 \\ 
        49 & 0.375 & 0.0005 & 0.002 	& 1650000 \\ 
        50 & 0.36 & 0.001 & 0.004 		& 1650000 \\ 
        51 & 0.073 & 0.0005 & 0.006 	& 6700000 \\ 
        52 & 0.055 & 0.0005 & 0.006 	& 6700000 \\ 
        53 & 0.063 & 0.0005 & 0.006 	& 6700000 \\ 
        54 & 0.041 & 0.0001 & 0.01 		& 6700000 \\ 
        55 & 0.02 & 0.0001 & 0.01 		& 6700000 \\ 
        56 & 0.073 & 0.0005 & 0.006 	& 6700000 \\ 
        57 & 0.063 & 0.0005 & 0.006 	& 6700000 \\ 
        58 & 0.055 & 0.0005 & 0.006 	& 6700000 \\ 
        59 & 0.094 & 0.0001 & 0.01 		& 6700000 \\ 
        60 & 0.015 & 0.0001 & 0.01 		& 6700000 \\ 
        61 & 0.039 & 0.0001 & 0.008 	& 1650000 \\ 
        62 & 0.13 & 0.0005 & 0.008 		& 1650000 \\ 
        63 & 0.34 & 0.0005 & 0.006 		& 4000000 \\ 
        64 & 0.035 & 0.001 & 0.004 		& 1650000 \\ 
        65 & 0.049 & 0.001 & 0.004 		& 1650000 \\ 
        66 & 0.36 & 0.001 & 0.004 		& 1650000 \\ 
        67 & 0.375 & 0.0005 & 0.002 	& 1650000 \\ 
        68 & 0.425 & 0.0005 & 0.002 	& 1650000 \\ 
        69 & 0.28 & 0.0005 & 0.003 		& 1650000 \\ 
        70 & 0.0167 & 0.0005 & 0.005 	& 1650000 \\ 
        71 & 0.008 & 0.0005 & 0.004 	& 1650000 \\ 
        72 & 0.0175 & 0.0005 & 0.005 	& 1650000 \\ 
        73 & 0.025 & 0.001 & 0.005 		& 6700000 \\ 
        74 & 0.027 & 0.001 & 0.005 		& 6700000 \\ 
        75 & 0.025 & 0.001 & 0.005 		& 6700000 \\ 
        76 & 0.047 & 0.001 & 0.005 		& 6700000 \\ 
        77 & 0.054 & 0.001 & 0.005 		& 6700000 \\ 
        78 & 0.01 & 0.001 & 0.005 		& 6700000 \\ 
        79 & 0.034 & 0.001 & 0.005 		& 6700000 \\ 
        80 & 0.038 & 0.001 & 0.005 		& 6700000 \\ 
        81 & 0.049 & 0.001 & 0.01 		& 6700000 \\ 
        82 & 0.016 & 0.001 & 0.01 		& 6700000 \\ 
        83 & 0.095 & 0.0004 & 0.0015 	& 1650000 \\ 
        84 & 0.053 & 0.001 & 0.01 		& 6700000 \\ 
        85 & 0.35 & 0.001 & 0.006 		& 1650000 \\ 
        86 & 0.021 & 0.001 & 0.004 		& 1650000 \\ 
        87 & 0.345 & 0.0005 & 0.003 	& 1650000 \\ 
        88 & 0.325 & 0.001 & 0.004 		& 1650000 \\ 
        89 & 0.14 & 0.0005 & 0.0015 	& 1650000 \\ 
        90 & 0.11 & 0.0005 & 0.0065 	& 6700000 \\ 
        91 & 0.045 & 0.0005 & 0.0065 	& 6700000 \\ 
        92 & 0.06 & 0.0005 & 0.006 		& 6700000 \\ 
        93 & 0.029 & 0.0005 & 0.006 	& 6700000 \\ 
        94 & 0.054 & 0.0005 & 0.004 	& 6700000 \\ 
        95 & 0.036 & 0.0005 & 0.004 	& 6700000 \\ 
\end{tabular}
\end{table}
\newpage

\subsection*{Peripheral Resistance Measurement}
The peripheral resistances were measured by the definition of different regional groups (see table \ref{peripheralGroups}). The peripheral resistance, $R_p$, of each group was determined by the volume difference between the steady state and running system. Volume integration was done by disconnection of the reservoirs and determination of the fluid amount per time. Only the corresponding arteries in the defined group were connected to the arterial network of the simulator by closing all 3-way valves to other arteries. Given the volume and pressure difference, the peripheral resistance for each group was calculated by using equation \ref{eqn:R_p}.

The peripheral resistance is build by different elements (see figure \ref{PeripheralElements}), which all possess static values, except the resistance of the small inner tubes is varying according to their length. In table \ref{TableR} the length of the small inner tubes to the corresponding boundary node ID is shown.

\begin{table}[H]
\caption{Values for the length, $l_p$, of the peripheral resistance elements ($d = 1\: mm$) for each boundary node, ID, referring to the $R_p$ defined in figure \ref{schematic}. }
\vspace{0.5cm}
\label{TableR} 
\begin{tabular}{cc}
\hline
Node ID &   $l_p$ (mm)  \\ 
\hline 
2  &  19.8  \\
3  &  19.4  \\
12 &  7  	\\
13 &  8.3  	\\
19 & 2.4  	\\
21 & 2.7  	\\
33 & 19.5  	\\
43 & 13  	\\
47 & 8.3  	\\
49 & 17.6  	\\
50 & 17.2  	\\
52 &19.4  	\\
53 & 19.4  	\\
57 & 19.4  	\\
58 & 19.5  	\\
66 & 17.6  	\\
67 & 17.2  	\\
68 & 4.7  	\\
69 & 13.3  	\\
70 & 8.3  	\\
71 & 19.4  	\\
72 & 8.2 	\\
76 & 8.4  	\\
79 & 5.4  	\\
80 & 8.4  	\\
87 & 8.9  	\\
88 & 7  	\\
93 & 2.3  	\\
95 & 2.1  	\\
7  &  / 	 \\
89 &   / 	 \\
14 &   / 	 \\
83 &   / 	 \\
\hline
\end{tabular}
\end{table}

\end{document}